\documentclass[aip,pof,preprint]{revtex4-1}
\usepackage{graphicx}
\usepackage{amssymb}
\usepackage{amsmath}
\usepackage{amsfonts}

\begin{document}
\title{Transport of Brownian particles confined to a weakly corrugated ``channel''}
\author{Xinli Wang}
\author{German Drazer}
\affiliation{Department of Chemical and Biomolecular Engineering \\ The Johns Hopkins University, Baltimore, MD 21218 USA}
\date{28 April 2008}

\begin{abstract}
We investigate the average velocity of Brownian particles driven by a constant external force when
constrained to move in two-dimensional, weakly-corrugated channels. We consider both the
geometric confinement of the particles between solid walls as well as the soft confinement
induced by a periodic potential.
Using perturbation methods we show that the leading order correction to the marginal
probability distribution of particles in the case of soft confinement is equal to that obtained
in the case of geometric confinement, provided that the (configuration) integral over the cross-section of the
confining potential is equal to the width of the solid channel.
We then calculate the probability distribution and average velocity in the case of a sinusoidal variation in the width of the channels.
The reduction on the average velocity is larger in the case of soft channels at small P\'eclet
numbers and for relatively narrow channels and the opposite is true at large P\'eclet numbers and for
wide channels. In the limit of large P\'eclet numbers the convergence to bulk velocity is faster in the
case of soft channels. The leading order correction
to the average velocity and marginal probability distribution agree well with Brownian Dynamics simulations for the two
types of confinement and over a wide range of P\'eclet numbers.
\end{abstract}

\maketitle

%%%%%%%%%%%%%%%%%%%%%%%%%%%%%%%%%%%%%%%%%%%%%%%%%%
%%%%%%%%%%%%%%%%%%%%%%%%%%%%%%%%%%%%%%%
\section{Introduction}
%%%%%%%%%%%%%%%%%%%%%%%%%%%%%%%%%%%%%%%%%%%%%%%%%%%
%%%%%%%%%%%%%%%%%%%%%%%%%%%%%%%%%%%%%%
The transport of Brownian particles confined to a channel with periodically varying cross section is relevant
to a broad range of problems,  from tracer dispersion in porous media to diffusion across entropy barriers in
biological systems.\cite{BuradaHMST09,Makhnovskii2010}
In addition, the transport of suspended species under geometric confinement is ubiquitous in the rapidly
growing field of microfluidics. Various separation microdevices, for example, are based on the effect that
heterogeneous microstructures and geometric confinement have on the average velocity of suspended
particles.\cite{Pamme07}

\begin{figure}[!ht]
\centering
\includegraphics[width=4in]{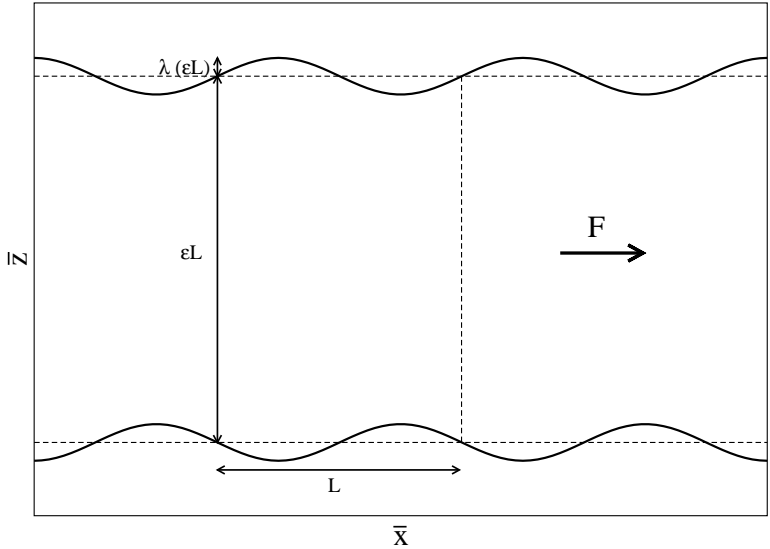}
\caption{Schematic diagram of the channel. The upper and lower ``boundaries'' of the channel correspond to the
solid walls in the case of geometric confinement or to two representative equipotential lines in the case of soft
confinement. }
\label{fig:channel}
\end{figure}

A number of studies have investigated the case of narrow channels, in which the characteristic length scale
of the cross-section is much smaller than the length of a single period along the channel. In the schematic of
 a two-dimensional wavy-wall channel shown in Fig. \ref{fig:channel} a narrow channel would correspond
 to $\epsilon \ll 1$. In a narrow channel, the description of diffusive transport in the absence of an external
 force can be simplified by reducing the dimensionality of the problem via the Fick-Jacobs approximation,
in which the motion in the cross section is transformed into an entropic {\it barrier} to longitudinal transport. \cite{Jacobs67,Zwanzig92,RegueraR01,KalinayP05,BuradaHMST09}
In the presence of an external field a relatively simple extension of the Fick-Jacobs approximation has been
used to calculate the average velocity of Brownian particles.\cite{RegueraSBRRH06, BuradaSRRH07, BuradaHMST09}
Alternatively, Laachi {\it et al.} \cite{LaachiKYD07} used the standard long-wave asymptotic perturbation
analysis and obtained analogous results for the average velocity to leading order in $\epsilon$.
In fact, we have shown that the leading order term in a perturbation analysis is equivalent to a direct extension
of the Fick-Jacobs approximation to the case of biased transport.\cite{WangD09}
In addition, and motivated by recently proposed partition-induced separation devices in which suspended species
are transported above a patterned surface,\cite{Bernate10}
we also considered the transport of Brownian particles confined by a relatively narrow channel in the potential
energy landscape.\cite{WangD09}
Remarkably, we found that the confining potential has identical effects on the transport of Brownian particles to
those induced by a solid channel, to leading order in the aspect ratio $\epsilon$.

In this work, we study the transport of Brownian particles through a two-dimensional wavy-wall channel in the limit
of small variations in the width of the channel, that is $\lambda \ll 1$ in the schematic shown in Fig. \ref{fig:channel}.
On the other hand, we shall assume that, in general, $\epsilon\sim O(1)$. Following our previous work we consider
both the geometric confinement by solid walls as well as the {\it soft} confinement induced by a periodic energy landscape.
This means that the channel boundaries shown in Fig. \ref{fig:channel} should be interpreted either as the solid
walls or as two representative equipotential lines. In both cases we investigate the effect that a small perturbation on
an otherwise flat channel has on the average velocity of the particles.

%%%%%%%%%%%%%%%%%%%%%%%%%%%%%%%%%%%%%%%%%%%%%%%%%%%%%%%%%%%%%%%%%%%%%%%%%%%%%%%%%%%%%%%%%%%
\section{Model description and previous results}
\label{chp:model}
%%%%%%%%%%%%%%%%%%%%%%%%%%%%%%%%%%%%%%%%%%%%%%%%%%%%%%%%%%%%%%%%%%%%%%%%%%%%%%%%%%%%%%%%%%%
Consider the transport of Brownian particles in the two different types of wavy channels discussed above.
In one case, the particles are confined between two solid walls. In the other case, the particles move in a
potential landscape that confines them in the direction perpendicular to the plane of motion. Specifically,
in the case of geometric confinement we consider a two-dimensional, symmetric channel with
half-width $\bar h(\bar x, \bar z) \equiv \bar h(\bar x)$ in the $\bar z$-direction and periodic in the $\bar x$-direction, $\bar h(\bar x+L)=\bar h(\bar x)$.
In the case of soft confinement we consider a periodic, two-dimensional potential, $\overline V(\bar x+L,\bar z)=\overline V(\bar x, \bar z)$,
that confines the particles in the $\bar z$-direction, i. e. ${\overline V}(\bar x,\bar z) \to +\infty$ for $\bar z \to \pm \infty$.
In both cases, in the limit of negligible inertia, the motion of Brownian particles can be described by the
Smoluchowski equation for the probability density $\overline P(\bar x,\bar z,t)$,
\begin{equation}
\label{eqn-smol}
\frac{\partial \overline P}{\partial t}+\nabla\cdot {\bf \overline J}=\delta(\bar x,\bar z)\delta(t).
\end{equation}
The probability flux, ${\bf \overline J}(\bar x,\bar z,t)$, is given by
\begin{equation}
\label{J} 				%% LABEL EQUATION: Definition of Flux %%
{\bf \overline J} = \frac{1}{\eta}\left(F\overline P - \frac{\partial \overline V}{\partial \bar x} \overline P - k_{B}T\frac{\partial \overline P}{\partial \bar x}\right)\vec{i} + \frac{1}{\eta}\left(-\frac{\partial \overline V}{\partial\bar z} \overline P-k_{B}T\frac{\partial \overline P}{\partial \bar z}\right)\vec{k},
\end{equation}
where $F$ is a uniform external force driving the particles in the $\bar x$-direction,
$\eta$ is the viscous friction coefficient, $k_B$ is the Boltzmann constant,
$T$ is the absolute temperature, and we have used the Stokes-Einstein equation to write the diffusion coefficient as a function of
$\eta$, $D=k_{B}T/\eta$. Let us mention that in the case of geometric confinement we shall assume that there is no potential field
present in the system, that is $\overline V(\bar x,\bar z)=0$.

In order to obtain the asymptotic distribution of particles, relative to a single period of the channel, we first
introduce the reduced probability density and the reduced probability current (see Refs. \onlinecite{Reimann02, LiD07}
for a more detailed discussion and Ref. \onlinecite{BrennerE93} for an analogous approach based in macrotransport theory),
\begin{eqnarray}
\label{reduced} 		%% LABEL EQUATION: Reduced Probability %%
\tilde P(\bar x,\bar z,t)= \sum_{n_x=-\infty}^{+\infty} \overline P(\bar x+n_x L,\bar z,t), \\
\mathbf{\tilde J}(\bar x,\bar z,t)= \sum_{n_x=-\infty}^{+\infty}\mathbf{\overline J}(\bar x+n_xL,\bar z,t).
\end{eqnarray}

The reduced probability is then obtained by solving Eq. (\ref{eqn-smol}) with periodic boundary conditions in $\bar x$,  $\tilde P(\bar x,\bar z,t)=\tilde P(\bar x+L,\bar z,t)$,
the no-flux condition in the $\bar z$-direction, and the normalization condition over a unit cell.
The no-flux condition imposed by the confinement depends on the type of channel.
In the case of geometric confinement, the condition at the solid walls is
\begin{equation}
{\bf \tilde{J}}\cdot\vec{n}={\tilde J}^x \; \bar h'(\bar x) \pm {\tilde J}^z = 0
\label{eqn:solid-flux} 	%% LABEL EQUATION: Flux at the solid walls %%
\end{equation}
where $\vec{n}=\bar h'(\bar x) \vec{i} \pm \vec{k}$ is a vector normal to the channel top and bottom walls defined by $\bar z=\mp \bar h(\bar x)$, respectively.
In the case of soft confinement, the no-flux condition at the solid boundary is replaced
by a far-field condition in ${\tilde J}^z$, i.e. a vanishingly small probability density (and flux) due to the confining potential,
\begin{equation}
\tilde{J}_{\infty}^{z} = \frac{1}{\eta}\left(-\frac{\partial {\overline{V}}}{\partial \bar z} \overline P - k_{B}T \frac{\partial \overline P}{\partial \bar z}\right) \xrightarrow[{\bar z \to \pm \infty}]{}  0.
\label{eqn:flux} 			%% LABEL EQUATION: Asymptotic Flux %%
\end{equation}

In what follows we simplify the analysis by using dimensionless variables, with $L$ as the characteristic length along the channel and,
to be consistent with our previous analysis in the case of narrow channels, we choose $\epsilon L$ as the  characteristic length scale in the cross-section,
where in this case $\epsilon$ is not necessarily small. Then, we introduce the dimensionless variables $x=\bar x /L$ and $z=\bar z /(\epsilon L)$, as well as
the dimensionless potential, $V = \overline{V}/(k_B T)$, and probability density
$P(x,z,t)=\epsilon L^2 \tilde{P}(\bar x, \bar z, t)$. The governing equation for the asymptotic probability distribution, $P_{\infty}(x,z)$, becomes
\begin{equation}
\epsilon^{2}\frac{\partial}{\partial x}\left[\left(\textrm{Pe}-\frac{\partial V}{\partial x}\right)P_{\infty}-\frac{\partial P_{\infty}}{\partial x}\right] + \frac{\partial}{\partial z}\left[-\frac{\partial V}{\partial z}P_{\infty}-\frac{\partial P_{\infty}}{\partial z}\right] = 0,
\label{eqn:governing}
\end{equation}
where the  P\'eclet number is defined as $\textrm{Pe}=FL/k_B T$. The periodic boundary condition in dimensionless form is
$P_{\infty}(x+1,z)=P_{\infty}(x,z)$, and the normalization condition becomes
\begin{equation}
\iint_\Omega P_\infty dx \, dz=1,
\label{eqn:norm}		%% LABEL EQUATION: Normalization  %%
\end{equation}
where the domain $\Omega$ is given by $\{(x,z):\;0\leq x\leq 1, -\infty<z<\infty\}$ for the case of soft confinement, or $\{(x,z): \; 0\leq x\leq 1, -h(x)\leq z\leq h(x)\}$ for the case of geometric confinement by solid walls.
Finally, the dimensionless no-flux condition in the case of soft confinement is essentially the same as in Eq. (\ref{eqn:flux}), that is $J^z_{\infty} \to 0$ for $z \to \pm \infty$.
On the other hand, the no-flux condition at the solid boundary becomes
\begin{equation}
{\bf \tilde{J}}\cdot\vec{n}= \epsilon^2 J^x \; h'(x) \pm  J^z = 0,
\label{eqn:solid-flux} 	%% LABEL EQUATION: Flux at the solid walls NONDIMENSIONAL %%
\end{equation}
due to the different characteristic scales used to nondimensionalize $x$ and $z$.

After the asymptotic solution for the reduced probability distribution is determined we can obtain the average
velocity along the channel by applying macrotransport theory,\cite{BrennerE93}
\begin{equation}
\langle v\rangle=\iint_\Omega J_\infty^{x} \, dx \, dz
=\iint_\Omega \left[\left(\textrm{Pe}-\frac{\partial V}{\partial x}\right)P_{\infty}-\frac{\partial P_{\infty}}{\partial x}\right] dx \, dz.
\label{eqn:v}			%% LABEL EQUATION: Average Velocity %%
\end{equation}

Alternatively, the motion of Brownian particles in a viscous solvent in the limit of vanishingly small inertia can also be described by the overdamped Langevin equations,
\begin{equation}
\eta\frac{dx}{dt}=F-\frac{\partial V}{\partial x}+\sqrt{\eta k_{B}T}\zeta_x(t),
\end{equation}
and
\begin{equation}
\eta\frac{dz}{dt}=-\frac{\partial V}{\partial z}+\sqrt{\eta k_{B}T}\zeta_z(t),
\end{equation}
where {\boldmath{$\zeta$}}$(t)$ is a zero-mean, Gaussian white noise, with independent components
satisfying the fluctuation-dissipation theorem, $\langle\zeta_{i}(t_{1})\zeta_{j}(t_{2})\rangle=2\delta_{ij}\delta(t_{1}-t_{2})$.
In this case, the average velocity along the channel can be evaluated from an ensemble average of independent trajectories,
\begin{equation}
\langle v\rangle=\lim_{t\to\infty} \frac{\left\langle x(t) \right \rangle}{t}.
\end{equation}

%%%%%%%%%%%%%%%%%%%%%%%%%%%%%%%%%%%%%%%%%%%%%%%%%%%%%%%%%%%%%%%%%%%%%%%%%%%%%%%%%%%%%%%%%%%
\subsection{Asymptotic analysis in the narrow channel approximation}
\label{recap}
%%%%%%%%%%%%%%%%%%%%%%%%%%%%%%%%%%%%%%%%%%%%%%%%%%%%%%%%%%%%%%%%%%%%%%%%%%%%%%%%%%%%%%%%%%%
In Ref. \onlinecite{WangD09} we showed that, in the case of narrow channels ($\epsilon \ll 1$),  the leading order solution of the reduced probability
is equivalent to the probability distribution that is obtained from the extension of the Fick-Jacobs approximation to the case of biased
transport. Moreover, we showed that  the geometric confinement between solid walls has the same effect on the transport properties of Brownian particles
as the confinement by a soft potential, to leading order in $\epsilon$. Specifically, the governing equations for both geometric and soft confinement cases are
the same, provided that the channel width, $w(x)=2h(x)$, is
equal to the following integral over the cross section,
\begin{equation}
\label{eqn:i}			%% LABEL EQUATION: Average Velocity %%
I(x)=\int_{-\infty}^{\infty}e^{-V(x,z)}dz.
\end{equation}
Note that, in the original work by Zwanzig, \cite{Zwanzig92} this integral was associated with a free energy $A(x)$ given by $I(x)=\exp(-A(x))$.
It can also be considered as a configuration integral over all possible states (vertical positions of the particle) at a given position along the channel.
The case of solid walls can be modeled as an infinite square-well, in which the potential is zero inside the channel and infinite outside. In this case
it is immediate that $I(x)=w(x)$ and the free energy is entropic.

If the condition described above is satisfied the leading order correction to the average velocity is also the same in both cases,
\begin{equation}
\label{v0}
\left<v_{FJ} \right>=(1-e^{-\textrm{Pe}})\left[\int_{0}^{1}dxe^{\textrm{Pe}\,x}I(x)\int_{x}^{x+1}
\frac{d\zeta}{e^{\textrm{Pe}\,\zeta}I(\zeta)}\right]^{-1}.
\end{equation}

%%%%%%%%%%%%%%%%%%%%%%%%%%%%%%%%%%%%%%%%%%%%%%%%%%%%%%%%%%%%%%%%%%%%%%%%%%%%%%%%%%%%%%%%%%%
%%%%%%%%%%%%%%%%%%%%%%%%%%%%%%%%%%%%%%%%%%%%%%%%%%%%%%%%%%%%%%%%%%%%%%%%%%%%%%%%%%%%%%%%%%%
\section{Asymptotic analysis in the small perturbation approximation}
\label{chp:asymptotic}
%%%%%%%%%%%%%%%%%%%%%%%%%%%%%%%%%%%%%%%%%%%%%%%%%%%%%%%%%%%%%%%%%%%%%%%%%%%%%%%%%%%%%%%%%%%
%%%%%%%%%%%%%%%%%%%%%%%%%%%%%%%%%%%%%%%%%%%%%%%%%%%%%%%%%%%%%%%%%%%%%%%%%%%%%%%%%%%%%%%%%%%
\subsection{Soft confinement by a periodic potential}
\label{chp:soft}
%%%%%%%%%%%%%%%%%%%%%%%%%%%%%%%%%%%%%%%%%%%%%%%%%%%%%%%%%%%%
We consider the transport of Brownian particles confined to a channel by a potential of the form
\begin{equation}
V(x,z)=V_0(z)(1+\lambda V_1(x)+\lambda^2 V_2(x)+\cdots).
\end{equation}
The potential $V_0(z)$, by itself, would confine the particles to a straight channel, assuming that $V_0(z) \to \infty$ for $z \to \pm \infty$.
The periodic functions $V_n(x)$ (for $n= \rm{1, 2, 3} \cdots$) therefore act as a perturbation to the channel width,
where the magnitude of the perturbation is determined by $\lambda$.
Here, we analyze the case of small perturbations, that is $\lambda\ll 1$. 
Therefore, we look for a solution to Eq. (\ref{eqn:governing}) in the form of a regular perturbation expansion for the asymptotic probability distribution,
\begin{equation}
P_\infty(x,z) \sim p_0(x,z)+\lambda p_1(x,z) + \lambda^2 p_2(x,z)+\cdots,
\label{eqn:P_series}
\end{equation}
where each function $p_n(x,y)$ ($n= \rm{0, 1, 2, 3} \cdots$) is periodic in $x$, $p_{n}\left(x=0,z\right) = p_{n}\left(x=1,z\right)$.
Analogously, we write a regular perturbation expansion for the probability flux,
\begin{equation}
{\bf J}_{\infty}(x,z) \sim {\bf J}_{0}(x,z)+\lambda{\bf J}_{1}(x,z)+\lambda^{2}{\bf J}_{2}(x,z)+\cdots.
\end{equation}
Let us also write the perturbation expansion of the configuration integral in Eq. (\ref{eqn:i}) explicitly as
\begin{equation}
\label{configuration}
\begin{split}
I(x)&=I_0 \left(1+\lambda I_1+O(\lambda^2) \right) \\
&=\int_{-\infty}^{\infty}e^{-V_0(z)}dz \left( 1-\lambda V_1(x) \overline V_0+O(\lambda^2) \right),
\end{split}
\end{equation}
where
\begin{equation} 
\overline V_0 = \frac{ \int_{-\infty}^{\infty}V_0(z) e^{-V_0(z)}dz}{\int_{-\infty}^{\infty}e^{-V_0(z)}dz}
\end{equation}

Substituting the perturbation expansion for the probability density into the governing equation, Eq. (\ref{eqn:governing}),
and equating like powers of $\lambda$ we obtain a hierarchy of coupled equations for the different functions $p_n(x,z)$.
The equations governing the basic solution, $p_0(x,z)$, and the perturbation functions $p_1(x,z)$ and $p_2(x,z)$ are:
\begin{equation}
\epsilon^2\frac{\partial}{\partial x}\left(\textrm{Pe}\;p_0-\frac{\partial p_0}{\partial x}\right)
+\frac{\partial}{\partial z}\left(-\frac{d V_0}{d z}p_0-\frac{\partial p_0}{\partial z}\right)=0,
\label{eqn:governing_0}
\end{equation}
\begin{equation}
\epsilon^2\frac{\partial}{\partial x}\left(\textrm{Pe}\;p_1-V_0 \frac{d V_1}{dx} p_0- \frac{\partial p_1}{\partial x}\right)
+\frac{\partial}{\partial z}\left(-\frac{dV_0}{dz}\left(p_1 + V_1 p_0 \right)-\frac{\partial p_1}{\partial z}\right)=0,
\label{eqn:governing_1}
\end{equation}
and
\begin{equation}
\begin{split}
\epsilon^2\frac{\partial}{\partial x}\left[\textrm{Pe}\;p_2-V_0 \left(\frac{dV_2}{dx}p_0+\frac{dV_1}{dx}p_1\right)-\frac{\partial p_2}{\partial x}\right]\\
+\frac{\partial}{\partial z}\left(-\frac{dV_0}{dz}\left(p_2+V_1 p_1+V_2 p_0\right)-\frac{\partial p_2}{\partial z}\right)=0,
\label{eqn:governing_2}
\end{split}
\end{equation}
which correspond to the $O(1)$, $O(\lambda)$, and $O(\lambda^2)$ terms, respectively.
Analogously, replacing the regular expansion for the flux in Eq. (\ref{eqn:flux}) we obtain,
\begin{equation}
J_{n}^{z}\left(x,z \to \pm\infty\right)=0, \qquad {\rm for} \quad n=\rm{0,1,2} \cdots.
\end{equation}
The normalization condition for the probability density corresponds to a basic solution $p_0(x,z)$ that is normalized to unity,
\begin{equation}
\left< p_{0} \right> = \int_0^1dx\int_{-\infty}^\infty dz ~p_0=1,
\end{equation}
and to the perturbation functions having zero average $\langle p_n\rangle=0$ for $n = 1, 2, 3 \cdots$.

In order to calculate the average velocity, we first simplify Eq. (\ref{eqn:v}) using the normalization condition and the periodicity
of the probability distribution,
\begin{equation}
\left< v \right> = \textrm{Pe} - \int_0^1dx\int_{-\infty}^\infty dz ~ \frac{\partial V}{\partial x} P_{\infty}.
\end{equation}

Replacing the expansion for the probability distribution into the equation above we obtain the general expansion for the average velocity to order $\lambda^2$,
\begin{eqnarray}
\label{vexpansion0}
v_0&=&\textrm{Pe}, \\
\label{vexpansion1}
v_1&=& -\int_0^1dx\int_{-\infty}^\infty dz ~ V_0 \frac{d V_1}{d x}  ~ p_0,\\
\label{vexpansion2}
v_2&=& -\int_0^1dx\int_{-\infty}^\infty dz ~ V_0 \left(\frac{d V_2}{d x}p_0+\frac{d V_1}{d x}p_1\right).
\end{eqnarray}

The basic solution is completely determined by  Eq. (\ref{eqn:governing_0}) and
the normalization condition, and we obtain:
\begin{equation}
p_0=\frac{1}{I_0}e^{-V_0(z)},
\label{eqn:p0_general}
\end{equation}
where $I_0$ was defined in Eq. (\ref{configuration}). Clearly, this basic solution is independent of $x$ and corresponds to the confinement by a straight channel in the absence of perturbations.
Therefore, it is easy to see from Eq. (\ref{vexpansion1}) that $v_1=0$ due to the periodicity of the perturbation potential $V_1(x)$.
Then, in order to obtain the leading order correction to the average velocity, we first need to determine the leading order correction to the probability density,
that is $p_1(x,z)$. The eigenvalue problem associated with Eq. (\ref{eqn:governing_1}) and the representation of the general solution in terms of eigenfunctions
are discussed in Appendix \ref{appendixEigenfunctions} and, in the next section, we will present the exact solution for the case of a parabolic potential
confining the particles in the $z$-direction.
Here, we limit the general discussion to the simpler problem of
finding the correction to the marginal probability density at $O(\lambda)$, that is
\begin{equation}
\bar  p_1(x)=\int_{-\infty}^\infty p_1(x,z)dz.
\end{equation}
The governing equation for $\bar p_1(x,z)$ is obtained by integrating Eq. (\ref{eqn:governing_1}) over the cross section, which yields
\begin{equation}
\label{eqn:p1bar_governing}		%% LABEL EQUATION: p1 %%
\frac{d}{dx}\left(\textrm{Pe}\bar p_1+\frac{d I_1}{dx}-\frac{d\bar p_1}{dx}\right)=0,
\end{equation}
where $I_1(x)=-V_1(x) \overline V_0$ was defined in Eq. (\ref{configuration}).
Integrating Eq. (\ref{eqn:p1bar_governing}), taking into account that $v_1=0$, and enforcing the normalization condition
we obtain
\begin{equation}
\label{p1bar}			%% LABEL solution to EQUATION: p1 %%
\bar p_1(x)=\frac{e^{\textrm{Pe}x}}{e^{-\textrm{Pe}}-1} \int_x^{x+1} e^{-\textrm{Pe}\,\zeta}
\,\frac{dI_1}{d\zeta}\,d\zeta.
\end{equation}

%%%%%%%%%%%%%%%%%%%%%%%%%%%%%%%%%%%%%%%%%%%%%%%%%%%%%%%%%%%%
\subsubsection{Particles confined by a parabolic potential}
\label{chp:parabolic}
%%%%%%%%%%%%%%%%%%%%%%%%%%%%%%%%%%%%%%%%%%%%%%%%%%%%%%%%%%%%
In order to explicitly calculate the leading order correction to the average velocity,
we consider the transport of Brownian particles confined to a channel by a parabolic potential of periodically fluctuating width.
Specifically,  the potential in non-dimensional variables is given by
\begin{equation}
V(x,z) = \frac{1}{2}\left( \frac{z}{\delta(x)}  \right)^{2},
\label{eqn:potential}
\end{equation}
where
\begin{equation}
\delta\left(x\right) =  \sqrt{\frac{2}{\pi}}\left(\frac{1}{2} +\lambda \sin(2\pi x)\right),
\end{equation}
The equilibrium probability is given by a Gaussian distribution with variance $\sigma(x)=\delta(x)$.
Let us mention that the factor $\sqrt{2}/\sqrt{\pi}$ included in the definition of $\delta(x)$ yields $I(x)=1+2\lambda \sin(2\pi x)$, which will be convenient when comparing the results
with the case of solid confinement. Clearly, in this case $I_0=1$ and $I_1(x)=2 \sin(2\pi x)$.

\begin{figure}[!ht]
\centering
\includegraphics[width=4in]{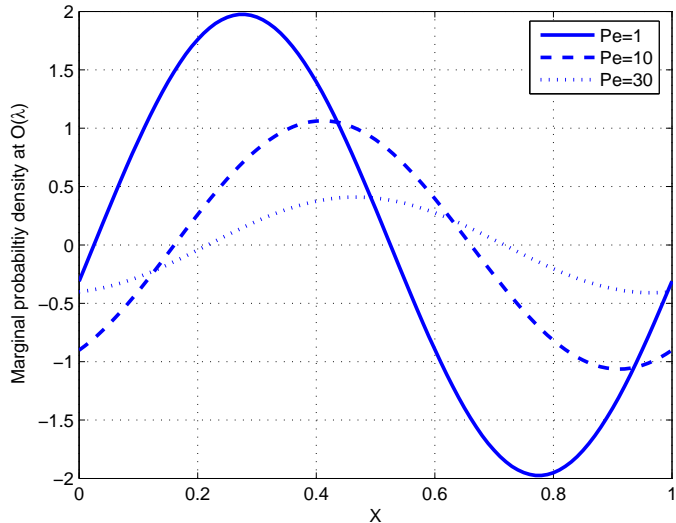}
\caption{Marginal probability density ${\bar p}_1(x)$ in the case of soft confinement by a parabolic potential, for different P\'eclet numbers.}
\label{fig:p1bar}
\end{figure}

The expansion of this potential is,
\begin{equation}
V(x,z) = \pi z^2 \left( 1 - 4 \lambda \sin(2\pi x) + 12 \lambda^2 \sin^2(2\pi x) + \cdots \right),
\end{equation}
and the basic solution is the Boltzmann distribution with $V_0(z)=\pi z^2$ as the potential energy,
\begin{equation}
p_0(z)=e^{-\pi z^2}.
\end{equation}
Therefore, the corresponding leading order marginal probability distribution is uniform.
The first correction to the marginal probability is obtained by direct integration of Eq. (\ref{p1bar}),
\begin{equation}
\label{eqn:p1bar}			%% LABEL EQUATION: marginal P1  %%
\bar p_1(x)=4\pi \left[ \frac{2\pi \sin2\pi x- \textrm{Pe}\cos2\pi x}{4\pi^2+\textrm{Pe}^2} \right].
\end{equation}
We can also write this marginal distribution in terms of a phase-shift $\phi_0$ with respect to the confining potential,
\begin{equation}
\bar p_1(x)=
2\cos\phi_0 \sin(2\pi x-\phi_0); \qquad \phi_0=\arctan\left( \frac{\textrm{Pe}}{2\pi} \right).
\end{equation}
Fig. \ref{fig:p1bar} shows the marginal probability for different P\'eclet numbers.
Interestingly, in the limit of small P\'eclet numbers we obtain ${\bar p}_1(x) \to 2\sin2\pi x$.
This is consistent with a uniform distribution across a solid
channel with a width given by $w(x)=I(x)=1+2\lambda\sin2\pi x$.
On the other hand, at large P\'eclet numbers, the distribution of particles integrated over
the cross-section becomes completely out of phase with respect to the variations of the channel width, that is $\phi_0 \to \pi/2$.
However, in this limit the magnitude of the correction to the basic solution vanishes.

As we discussed in the previous section, in order to obtain the leading order correction to the average
velocity, $v_2$, we need to calculate $p_1(x,z)$. The governing equation for $p_1(x,z)$ is obtained
from Eq. (\ref{eqn:governing_1}) by substituting the expressions for the
confining potential, $V_0=\pi z^2$, and the leading order perturbation, $V_1=-4\sin(2\pi x)$,
\begin{equation}
\label{eqn:parabolic1}
\epsilon^2\frac{\partial}{\partial x}\left(\textrm{Pe}p_1-\frac{\partial p_1}{\partial x}\right)+
\frac{\partial}{\partial z}\left(-2\pi z\,p_1-\frac{\partial p_1}{\partial z}\right)=  \left[(16 \epsilon^2\pi^3+16\pi^2) z^2 -8\pi\right] p_0 \sin(2\pi x).
\end{equation}

In Appendix \ref{appendixEigenfunctions} we show that,  in the case of a parabolic potential, $p_1(x,z)$ can be written in
terms of $p_0(z) H_0(\sqrt{\pi}z)$ and $p_0(z) H_2(\sqrt{\pi} z)$, where $H_m(x)$  are the Hermite polynomials,
$H_0(x)=1$ and $H_2(x)=(4x^2-2)$.
Let us note that the latter term, $2p_0(z) (2\pi z^2-1)$,  integrates
to zero over the cross section and thus it does not contribute to the marginal probability.
Therefore, we propose the following solution,
\begin{equation}
p_1(x,z)= \left[  2 \cos(\phi_0)  \sin(2\pi x-\phi_0) + (2\pi z^2-1) (a  \sin(2\pi x) + b \cos(2\pi x) ) \right]\,p_0(z),
\label{eqn:p1_coeff}
\end{equation}
where $a$ and $b$ are undetermined constants.
The proposed solution is clearly periodic and the normalization condition is also satisfied for arbitrary values of $a$ and $b$, by construction.
These two constants are determined by  substituting Eq. (\ref{eqn:p1_coeff}) into
Eq. (\ref{eqn:parabolic1}) and equating the coefficients of each of the orthogonal functions to zero.
The final solution is,
\begin{equation}
p_1(x,z)= 2\left[ \cos(\phi_0) \sin(2\pi x-\phi_0) +  \cos(\phi_1) (2\pi z^2-1) \sin(2\pi x-\phi_1) \right]\,p_0(z),
\end{equation}
where the phase shift $\phi_1$ is given by,
\begin{equation}
\phi_1 = \arctan\left(  \frac{\epsilon^2 \textrm{Pe}}{2(\pi\epsilon^2+1)} \right).
\end{equation}

Finally, the leading order correction to the average velocity, $v_2$, is obtained from Eq. (\ref{vexpansion2})
\begin{equation}
\begin{split}
v_2 &= -\int_0^1dx \int_{-\infty}^{\infty} dz \left( \pi z^2 \right) (-8\pi \cos(2\pi x))\,p_1(x,z) \\
&= -4\pi \left[ \cos(\phi_0) \sin(\phi_0) + 2 \cos(\phi_1) \sin(\phi_1) \right] \\
&= -8\pi\textrm{Pe}\left[ \frac{\pi}{4\pi^2+\textrm{Pe}^2} + \frac{2\epsilon^2(\pi\epsilon^2+1)}{\epsilon^4\textrm{Pe}^2+4(\pi\epsilon^2+1)^2} \right].
\end{split}
\end{equation}
Alternatively, we can calculate the effective mobility, defined as the ratio of the average velocity to the applied force. In dimensionless form, the effective mobility normalized by the mobility in bulk, $\mu_0=\eta^{-1}$,  is given by
\begin{equation}
\label{mobility-soft}
\frac{\mu^s_{\rm eff}}{\mu_0}=\frac{\langle v \rangle}{\textrm{Pe}}\approx 1-
8 \pi \lambda^2
\left[ \frac{\pi}{4\pi^2+\textrm{Pe}^2}
+ \frac{2\epsilon^2(\pi\epsilon^2+1)}{\epsilon^4\textrm{Pe}^2+4(\pi\epsilon^2+1)^2} \right] + O(\lambda^4)
\end{equation}
Finally, and for comparison with the case of geometric confinement, we calculate the limiting behavior at small P\'eclet numbers,
\begin{equation}
\label{small-pe-soft}
\lim_{\textrm{Pe}\to 0} \frac{\langle v \rangle}{\textrm{Pe}}=1-2\lambda^2\left[1+\frac{2\pi\epsilon^2}{\pi\epsilon^2+1}\right],
\end{equation}
and the asymptotic behavior at large P\'eclet numbers,
\begin{equation}
\label{large-pe-soft}
\frac{\langle v\rangle}{\textrm{Pe}}\sim 1-\lambda^2 \frac{8 \pi}{\epsilon^2\textrm{Pe}^2} \left( 2+3\pi \epsilon^2 \right).
\end{equation}

%%%%%%%%%%%%%%%%%%%%%%%%%%%%%%%%%%%%%%%%%%%%%%%%%%%%%%%%%%%%
\subsection{Geometric confinement by solid walls}
\label{chp:solid}
%%%%%%%%%%%%%%%%%%%%%%%%%%%%%%%%%%%%%%%%%%%%%%%%%%%%%%%%%%%%
Let us consider now the case of a solid channel that is periodic in $x$ and symmetric about the $xy$-plane.
The upper wall of the channel is given by $z=h(x)$, with
\begin{equation}
h(x)=\frac{1}{2}+\lambda g(x),
\end{equation}
where $\lambda$ is again assumed to be a small parameter. Note that this choice of channel width corresponds to $I_0=1$ in the 
case of soft-confinement.
The main objective is to determine the effect that such small oscillations in the width of channel have on the average velocity of the particle.
Therefore, we seek a solution to the asymptotic distribution in the form of a regular asymptotic expansion in
$\lambda$,
\begin{equation}
P_\infty(x,y) \sim \rho_0(x,z)+\lambda \rho_1(x,z) +\lambda^2 \rho_2(x,z) + \cdots.
\label{expansion-rho}
\end{equation}
Replacing the expansion above into the governing equation for the asymptotic probability, Eq. (\ref{eqn:governing}), and equating like powers of
$\lambda$, it becomes clear that in the absence of an external potential
the governing equations for successive powers of $\lambda$ are all the identical. Specifically, the general equation for $\rho_n$ is
\begin{equation}
\epsilon^2\frac{\partial}{\partial x}\left(\textrm{Pe}\rho_n-\frac{\partial \rho_n}{\partial x}\right)-\frac{\partial^2 \rho_n}{\partial z^2}=0.
\label{eqn:gov-solid}				%% LABEL EQUATION: Governing Equation%%
\end{equation}
Although the previous equation seems to indicate that the different functions $\rho_n(x,z)$ are independent, they are actually
coupled to each other through the boundary and normalization conditions. Therefore, we first use the domain perturbation method
\cite{Leal07, MalevichMA06} to transform the no-flux condition at the wavy walls, given by Eq. (\ref{eqn:solid-flux}), to
an asymptotically equivalent boundary condition at the two flat planes given by $z=\pm 1/2$. Analogously, Eq. (\ref{eqn:norm})
is transformed into an equivalent normalization condition using a Taylor series expansion for the
domain $\Omega$ (a more detailed discussion is presented in Appendix \ref{app_dp}).
Finally, applying the domain perturbation method to Eq. (\ref{eqn:v}) we shall obtain
the regular expansion for the average velocity of the particles.

Let us first rewrite Eq. (\ref{eqn:v}) replacing $P_\infty(x,z)$ by its regular expansion,
\begin{equation}
\langle v\rangle = \textrm{Pe} - \int_0^1dx\int_{-1/2-\lambda g(x)}^{1/2+
\lambda g(x)}\left(\frac{\partial \rho_0}{\partial x}+\lambda\frac{\partial \rho_1}{\partial x}
+\lambda^2\frac{\partial \rho_2}{\partial x}+O(\lambda^3)\right)dz.
\end{equation}
Before we expand the domain of integration in powers of $\lambda$, we note that the basic solution for the asymptotic distribution of particles is
uniform, that is $\rho_0=1$.
This is expected, given that in the present approximation the leading order term in $P_\infty(x,z)$
corresponds to the asymptotic distribution of particles between two flat plates.
As a result, $\rho_0$ does not contribute to the integral above.
Now, using a Taylor expansion about $\lambda=0$ for the domain of integration,
and taking into account the periodicity of $\rho_n$, i.e. $\int_0^1 (\partial \rho_n/\partial x) dx=0$,
we obtain the averge velocity to order $\lambda^2$,
\begin{equation}
\label{eqn:av_vel}				%% LABEL EQUATION: Average velocity %%
u_0=\textrm{Pe}; \qquad
u_1= 0; \qquad
u_2=-\int_0^1 \left[ \frac{\partial \rho_1}{\partial x} \bigg|_{z=1/2}+ \frac{\partial \rho_1}{\partial x} \bigg|_{z=-1/2}\right] g \, dx.
\end{equation}
Let us note that, in agreement with the case of soft confinement, the first order vanishes,
as expected given the symmetry of the problem. In fact, in both problems the geometry of the channel is not affected by a change $\lambda \to -\lambda$
and thus all the terms in the average velocity with odd powers
of $\lambda$ are zero.
Therefore, in order to calculate the leading order perturbation to the average velocity, $u_2$, we first need to determine $\rho_1(x,z)$.

Let us first  calculate the perturbation to the marginal probability distribution for an arbitrary function $g(x)$.
Integrating the regular expansion for $P_{\infty}(x,z)$ over the cross section and
expanding the integral in a Taylor series about $\lambda$, we obtain the following
expression for the order $\lambda$ contribution to the marginal probability,
\begin{equation}
\bar \rho_1(x)=2g(x)+\int_{-1/2}^{1/2}\rho_1(x,z) \, dz=2g(x) +\bar s_1(x),
\end{equation}
The governing equation for $\bar{s_1}(x)$ is then obtained by integrating Eq. (\ref{eqn:gov-solid}) (for $n=1$) with respect to $z$ from $-1/2$ to $1/2$
and using the corresponding no-flux boundary condition (see Appendix \ref{app_dp}),
\begin{equation}
\textrm{Pe}\frac{d \bar s_1}{d x}-\frac{d^2 \bar s_1}{d x^2}+2\textrm{Pe}\frac{dg}{dx}=0.
\label{eqn:avragewidth_solid1}
\end{equation}
In addition, $\bar{s_1}(x)$ inherits the periodicity and normalization conditions from $\rho_1(x,z)$.
The solution, after some manipulation, takes the form
\begin{equation}
\begin{split}
\bar s_1(x)
&=\frac{2\textrm{Pe} \, e^{\textrm{Pe}x}}{e^{-\textrm{Pe}}-1}\int_x^{x+1}e^{-\textrm{Pe}\zeta}g(\zeta)d\zeta\\
&=-2g(x)+\frac{2e^{\textrm{Pe}x}}{e^{-\textrm{Pe}}-1}\int_x^{x+1}e^{-\textrm{Pe}\zeta}\frac{dg}{d\zeta}d\zeta.
\end{split}
\end{equation}
Therefore, the leading order perturbation to the marginal probability distribution is given by,
\begin{equation}
\bar \rho_1(x) = \frac{2e^{\textrm{Pe}x}}{e^{-\textrm{Pe}}-1} \int_x^{x+1}e^{-\textrm{Pe}\zeta}\frac{dg}{d\zeta}d\zeta.
\end{equation}
This equation is analogous to Eq. (\ref{p1bar}), obtained in the case of soft confinement,
where the perturbation to the width of the solid channel, given by $2g(x)$, replaces the perturbation potential, $I_1(x)$. 
Moreover, taking into account that the unperturbed width of the channel considered here corresponds to $I_0=1$, then the
condition $I_1(x)=2 g(x)$ is equivalent to the condition found in the case of narrow channels and discussed in Sec. \ref{recap},
that is $I(x)=w(x)$.

\begin{figure}[!ht]
\centering
\includegraphics[width=4in]{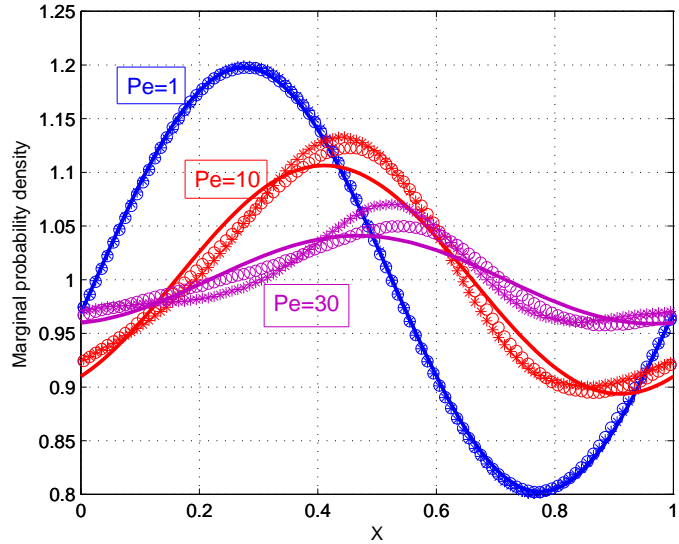}
\caption{Marginal probability density for different P\'eclet numbers. Solid lines refer to the asymptotic results up to $O(\lambda^2)$; circles$/$stars refer to the results from the Brownian dynamics simulation in the soft$/$solid confinement. Parameters used in the numerical simulation are $\lambda=0.1$ and $\epsilon=0.9$.}
\label{fig:pbar}
\end{figure}

In order to obtain an analytical expression for the leading correction to the average velocity
we consider the case $g(x)=\sin2\pi x$. This choice of $g(x)$ satisfies the condition derived above,
that is $I(x)=w(x)$, for the special case of soft confinement with a perturbation given by $V_1(x)=-4 \sin2\pi x$, discussed before in Sec. \ref{chp:parabolic}. Clearly, the leading order perturbation to the marginal probability is then the same as in the case of soft confinement,
\begin{equation}
\label{eqn:p1_bar_solid}
\bar \rho_1(x)=4\pi \left[ \frac{2\pi\sin2\pi x-\textrm{Pe}\cos2\pi x}{4\pi^2+\textrm{Pe}^2} \right].
\end{equation}
In Fig. \ref{fig:pbar} we compare the marginal probability density up to order $\lambda$ obtained from the asymptotic analysis with the results obtained from numerical simulations performed in both
soft and solid cases for different P\'eclet numbers, where solid lines refer to the asymptotic results up to $O(\lambda^2)$, and circles$/$stars refer to the results from the Brownian dynamics simulation in the soft$/$solid confinement. The agreement is excellent for P\'eclet numbers as large as $\textrm{Pe}=1$. Let us note that the
asymptotic analysis is valid for $\lambda \ll 1$ and assumes that $\lambda \textrm{Pe} \ll 1$ as well. For the cases shown in the figure this latter assumption is only valid if $\textrm{Pe} \ll 10$. The simulation
results show that, as expected, the effect of soft and geometric confinement on the distribution of particles is not the same for large P\'eclet numbers, 
e. g. for large values of the driving force.

The probability distribution $\rho_1(x,z)$  can be determined by proposing two base solutions of the form
\begin{equation}
\rho_1(x,z)=f_1(z)e^{2\pi ix} \quad\textrm{and}\quad g_1(z)e^{-2\pi i x}.
\label{eqn:rho1_solid}
\end{equation}

Then, substituting these general solutions into the governing equation and the equation for the
no-flux boundary condition we obtain (see details in appendix \ref{app_p1}),
\begin{equation}
\rho_1=2\mathbb{R}(A_1 \cosh(\alpha_1 z)e^{2\pi i x}),
\end{equation}
where $\mathbb{R}(\cdot)$ denotes the real part, and $A_1$ and $\alpha_1$ are the following complex numbers
\begin{equation}
A_1 =-\frac{\pi \epsilon^2\textrm{Pe}}{\alpha_1 \sinh(\alpha_1/2)} \quad\textrm{and}\quad \alpha_1^2 = \epsilon^2(2\pi i \textrm{Pe}+4\pi^2).
\end{equation}

Finally, we calculate the average velocity at $O(\lambda^2)$,
\begin{equation}
u_2=-\int_0^1 2\sin2\pi x\frac{\partial p_1}{\partial x}\bigg|_{z=1/2}dx=-4\pi^2\epsilon^2\textrm{Pe}\,\mathbb{R}\left(\frac{\coth(\alpha/2)}{\alpha}\right).
\end{equation}

\begin{figure}[!ht]
\centering
\includegraphics[width=4in]{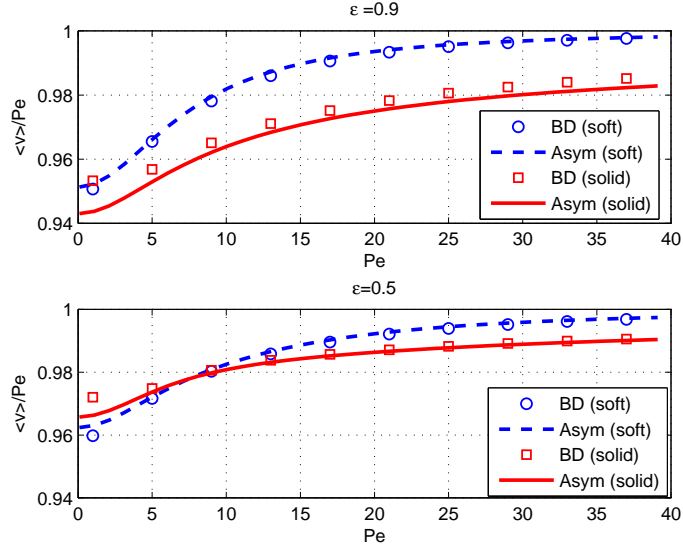}
\caption{(Color online) Effective mobility of Brownian particles in solid and soft channels as a function of the P\'eclet number.
The top plot corresponds to $\epsilon$=0.9 and the bottom one to $\epsilon$=0.5. In both cases we use $\lambda=0.1$.
The solid (dashed) line corresponds to the effective mobility in solid (soft) channels accurate to order $\lambda^2$.
The open squares (circles) correspond to the results of Brownian dynamics simulations in the case of geometric (soft) confinement.}
\label{fig:M_both}
\end{figure}

The corresponding effective mobility, normalized by the bulk mobility,  is given by
\begin{equation}
\label{mobility-wall}
\frac{\mu^w_{\rm eff}}{\mu_0} = \frac{\langle v\rangle}{\textrm{Pe}}\approx 1-4\lambda^2 \pi^2\epsilon^2\, \mathbb{R}\left(\frac{\coth(\alpha/2)}{\alpha}\right) + O(\lambda^4).
\end{equation}
Comparing the equation above with Eq. \ref{mobility-soft} we see that the two mobilities are different.
This is the first difference observed for the transport of Brownian particles between the cases of soft and
geometric confinement. Moreover, even the limiting behavior at small P\'eclet numbers is, in general,
different for the two types of channels.
The limit of the effective mobility at zero P\'eclet number is,
\begin{equation}
\lim_{\textrm{Pe}\to 0}\frac{\langle v\rangle}{\textrm{Pe}} =1-2\lambda^2\pi\epsilon\coth(\pi\epsilon),
\end{equation}
and the asymptotic behavior at large P\'eclet numbers is
\begin{equation}
\frac{\langle v\rangle}{\textrm{Pe}}\sim1-\lambda^2\frac{2\pi^2\epsilon}{\sqrt{\pi \textrm{Pe}}}.
\end{equation}
Let us note, however, that the effective mobilities are equal in the limit of narrow channels, as was already discussed in Sec. \ref{recap}.
In fact, taking the limit of $\epsilon \to 0$ we see that both effective mobilities tend to the same value, $\mu_{\rm{eff}}(\epsilon=0)=1-2\lambda^2$.

\begin{figure}[!ht]
\centering
\includegraphics[width=4in]{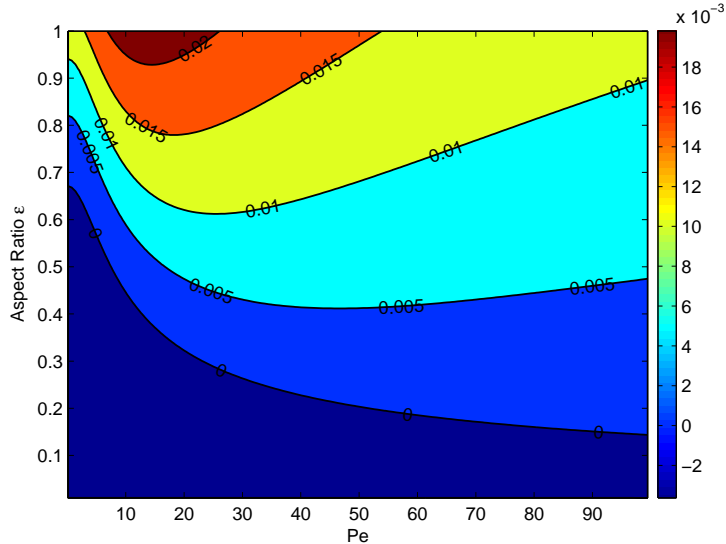}
\caption{(Color online) Contour plot of the difference in the effective mobility between the cases of soft and geometric confinement, $(\mu^s_{\rm eff} - \mu^w_{\rm eff})/\mu_0$.}
\label{fig:V_diff}
\end{figure}

A comparison between the effective mobility in a solid channel and that in a channel created by a confining potential  is presented
in Fig. \ref{fig:M_both}. The numerical results obtained from Brownian Dynamics simulations in both types of channels
are also shown in the figure.
The results from the numerical simulations agree very well  with the leading order corrections given by Eqs. (\ref{mobility-soft}) and (\ref{mobility-wall}),
over a wide range of P\'eclet numbers.
Moreover, in agreement with the asymptotic results obtained for large P\'eclet numbers, it is clear that the convergence of the effective mobility towards its bulk value
is slower in the case of geometric confinement and, as a result,  the mobility in solid channels is smaller at large P\'eclet numbers.
On the other hand, the effective mobility in the case of geometric confinement is typically larger than that in the case of soft confinement for small values of the aspect ratio $\epsilon$.
A detailed comparison of the mobilities in the cases of soft and geometric confinements as a function of both the aspect ratio $\epsilon$ and the P\'eclet number is
presented in Fig. \ref{fig:V_diff}.
We can see that  for aspect ratios above $\epsilon_c\approx0.67$, $\mu^s_{\rm eff} > \mu^w_{\rm eff}$ for all P\'eclet numbers.
On the other hand, for  smaller values of the aspect ratio,  the difference in mobilities goes from negative ($\mu^s_{\rm eff} < \mu^w_{\rm eff}$) at small P\'eclet numbers
to positive as the P\'eclet number increases. For arbitrary but fixed values of the P\'eclet number the mobility difference goes from negative at small aspect ratios to
positive as the aspect ratio increases.

%%%%%%%%%%%%%%%%%%%%%%%%%%%%%%%%%%%%%%%%%%%%%%%%%%%%%%%%%%%%
\section{Conclusion}
\label{chp:discussion}
%%%%%%%%%%%%%%%%%%%%%%%%%%%%%%%%%%%%%%%%%%%%%%%%%%%%%%%%%%%%
We have investigated the transport of Brownian particles confined by two-dimensional, weakly-corrugated channels.
We considered the cases of soft confinement, in which the particles move in a periodic potential, and of geometric confinement between solid walls.
In particular, we studied the effect that small variations in the width of the channels have on the average velocity of Brownian particles by means of
regular perturbation methods. We showed that the leading order correction to the marginal probability distribution in the case of soft confinement is
equivalent to that in the case of geometric confinement,
provided that the (configuration) integral of the soft-confinement potential over the cross-section is equal to the channel width in the case of solid walls.
The same condition was (previously) found in the case of narrow channels.
We then considered the specific case of sinusoidal variations in the channel width and calculated the leading
order correction to the probability distribution and to the average velocity of the particles.
We found that the reduction on the average velocity is different in the cases of soft and solid channels.
Interestingly, whereas the reduction in the velocity is larger in the case of soft channels at small P\'eclet numbers and relatively narrow channels,
the opposite is true at large P\'eclet numbers and for wide channels.
The asymptotic behavior at large P\'eclet numbers is also different, with a much faster convergence to the asymptotic velocity in the case
of soft confinement, with the perturbation decaying as $1/\textrm{Pe}^2$, compared to the case of geometric confinement in which the
perturbation to the velocity vanishes as  $1/\sqrt{\textrm{Pe}}$. We also performed Brownian Dynamics simulations for the two types of
channels and the numerical results agree well with the analytical results over a broad range of P\'eclet numbers.

\section*{Acknowledgments}
This material is partially based upon work supported by the National Science Foundation under grant No. CBET-07331023.
%%%%%%%%%%%%%%%%%%%%%%%%%%%%%%%%%%%%%%%%%%%%%%%%%%%%%%%%%%%%%%%%%%%%%%%%%%%%%%%%%%%%
\appendix
%%%%%%%%%%%%%%%%%%%%%%%%%%%%%%%%%%%%%%%%%%%%%%%%%%%%%%%%%%%%%%%%%%%%%%%%%%%%%%%%%%%%
\section{Eigenvalue problem and eigenfunction expansion}
\label{appendixEigenfunctions}
%%%%%%%%%%%%%%%%%%%%%%%%%%%%%%%%%%%%%%%%%%%%%%%%%%%%%%%%%%%%%%%%%%%%%%%%%%%%%%%%%%%%
We first transform the linear differential equation for $p_1(x,z)$ given by Eq. (\ref{eqn:governing_1}) into a self-adjoint linear differential equation for
${\tilde p}_1(x,z)= \exp{\left(-\frac{\textrm{Pe}\; x}{2}\right)} \exp\left(\frac{V_0(z)}{2}\right) p_1(x,z)$,
\begin{equation}
{\cal L}\left[{\tilde p}_1(x,z)\right]=\left( q(z)-\epsilon^2\frac{\partial^2}{\partial x^2}-\frac{\partial^2}{\partial z^2} \right){\tilde p}_1(x,z)=f(x,z),
\end{equation}
with
\begin{eqnarray}
q(z) &=& \epsilon^2\left(\frac{\textrm{Pe}}{2}\right)^2 + \left( \frac{1}{2} \frac{dV_0}{dz} \right)^2 - \frac{1}{2} \frac{d^2 V_0}{dz^2}, \\
f(x,z) &=& p_0(z)  e^{\left(-\frac{\textrm{Pe}\; x}{2}\right)} e^{\left(\frac{V_0(z)}{2}\right)}
\left[  \epsilon^2\frac{d^2 V_1}{dx^2} V_0 + V_1 \frac{d^2V_0}{dz^2} - V_1 \left( \frac{dV_0}{dz} \right)^2 \right].
\label{qandf}
\end{eqnarray}
${\cal L}$ is therefore a self-adjoint linear differential operator with an associated eigenvalue equation ${\cal L}\psi = \lambda \psi$ that is separable into two linear ordinary differential equations.
Specifically, writting $\psi = {\cal X}{\cal Z}$ we obtain
\begin{eqnarray}
\epsilon^2{\cal X}''(x) + \left\{ \alpha - \epsilon^2\left(\frac{Pe}{2}\right)^2 \right\} {\cal X}(x)&=&  0,   \\
{\cal Z}''(z)  + \left\{ \beta - \left(\frac{1}{2} \frac{dV_0}{dz} \right)^2 + \frac{1}{2} \frac{d^2 V_0}{dz^2} \right\} {\cal Z}(z)&=&  0,
\end{eqnarray}
and the eigenvalue equation for $\cal L$ holds with $\lambda=\alpha+\beta$.
The first equation is a standard Sturm-Liouville problem in the interval $(0,1)$ and the eigenfunctions
${\cal X}_n(x)$ are $\sin(n\pi x)$ and $\cos(n\pi x)$ with eigenvalues $\alpha_n=\epsilon^2(n^2\pi^2+\textrm{Pe}^2/4)$. The spectrum of the second eigenvalue problem is also discrete, assuming that
$q(z) \to \infty$ for $z \to \pm \infty$. We can therefore write the eigenfunctions and eigenvalues as ${\cal Z}_m(z)$ and $\beta_m$, respectively.

An interesting case is that of a quadratic potential, $V_0(z)=\pi z^2$, confining the particles in the $z$-direction. In this case, the solutions to the
second eigenvalue problem are ${\cal Z}_m(z)=\exp(-\pi z^2/2)\,H_m(\sqrt{\pi}z)$, where $H_m(x)$ are the Hermite polynomials,
and we can expand the general solution in known eigenfunctions,
\begin{equation}
{\tilde p}_1(x,z) = \sum_{n=1}^{\infty} \sum_{m=1}^{\infty} \left[c_{n,m} \sin(n\pi x) + d_{n,m} \cos(n\pi x)\right] ~ e^{\frac{-\pi z^2}{2}} ~ H_m \left( \sqrt{\pi}z \right),
\end{equation}
with
\begin{equation}
c_{n,m} =\frac{1}{2^n n!} \frac{1}{\left(n^2\pi^2+\frac{\textrm{Pe}^2}{4}\right)\epsilon^2+\pi m}
\int_0^1 dx \int_{-\infty}^{+\infty} dz~f(x,z) \sin(n\pi x) e^{\frac{-\pi z^2}{2}}\,H_m(\sqrt{\pi}z),
\end{equation}
\begin{equation}
d_{n,m} =\frac{1}{2^n n!} \frac{1}{\left(n^2\pi^2+\frac{\textrm{Pe}^2}{4}\right)\epsilon^2+\pi m}
\int_0^1 dx \int_{-\infty}^{+\infty} dz~f(x,z) \cos(n\pi x) e^{\frac{-\pi z^2}{2}}\,H_m(\sqrt{\pi}z),
\end{equation}
where $f(x,z)$ is obtained from Eq. (\ref{qandf}),
\begin{equation}
f(x,z)= e^{-\frac{\textrm{Pe}\; x}{2}}\,e^{- \frac{\pi\,z^2}{2}} \,
\left[ \epsilon^2 \pi z^2 \frac{d^2 V_1}{dx^2} +  2\pi\left(1-2\pi z^2 \right) V_1\right],
\end{equation}
 It is important to note that $f(x,z)$ can be written as a combination of ${\cal Z}_0(z)$ and ${\cal Z}_1(z)$ and, therefore, the only non-zero
 coefficients are $c_{n,0}$ and $c_{n,1}$. As a result, it is possible to write $p_1(x,z)$ as a linear combination of
 $p_0(z)$ and $p_0(z) (2\pi z^2-1)$.

%%%%%%%%%%%%%%%%%%%%%%%%%%%%%%%%%%%%%%%%%%%%%%%%%%%%%%%%%%%%%%%%%%%%%%%%%%%%%%%%%%%%
\section{Domain perturbation}
\label{app_dp}
%%%%%%%%%%%%%%%%%%%%%%%%%%%%%%%%%%%%%%%%%%%%%%%%%%%%%%%%%%%%%%%%%%%%%%%%%%%%%%%%%%%%

Here we apply the domain perturbation method \cite{Leal07, MalevichMA06} to transform the no-flux boundary condition at the curved wall defined by $z=h(x)=1/2+\lambda g(x)$ to an asymptotically equivalent
boundary condition applied at $z=1/2$. We start by substituting the regular perturbation expansion for the asymptotic probability density given by Eq. (\ref{expansion-rho})
into the no-flux boundary condition given by Eq. (\ref{eqn:solid-flux}),
\begin{equation}
\begin{split}
&\lambda\epsilon^2\frac{dg}{dx}
\left[\textrm{Pe}\left(\rho_0+\lambda \rho_1 \right)-\frac{\partial \rho_0}{\partial x}
-\lambda\frac{\partial \rho_1}{\partial x} \right]+ \frac{\partial \rho_0}{\partial z}+
\lambda\frac{\partial \rho_1}{\partial z}+\lambda^2\frac{\partial \rho_2}{\partial z}+ O(\lambda^3)=0,\\
&\textrm{at} \quad z=\frac{1}{2}+\lambda g(x).
\label{eqn:noflux_solid_P}
\end{split}
\end{equation}
Then, we approximate the functions $\rho_n(x,z)$ in the neighborhood of $z=1/2$ with a Taylor series in powers of $\lambda$,
\begin{equation}
\left. \rho_n \right|_{z=\frac{1}{2}+\lambda g(x)} = \rho_n |_{z=\frac{1}{2}} + \left. \frac{\partial \rho_n}{\partial z} \right|_{z=\frac{1}{2}}  \lambda g(x)+
\frac{1}{2} \left. \frac{\partial ^2 \rho_n}{\partial z^2}\right|_{z=\frac{1}{2}} \left( \lambda g(x) \right)^2+ O(\lambda ^3).
\label{eqn:pi_solid}
\end{equation}
Substituting these regular expansions into the no-flux boundary condition and equating like powers of $\lambda$ we obtain
the equivalent boundary conditions at $z=1/2$ for each power of $\lambda$,
\begin{equation}
\begin{split}
\label{eqn:asy_flux}
&\textrm{at} \quad O(1): \quad \frac{\partial \rho_0}{\partial z}=0 \quad \textrm{at}\quad z=\frac{1}{2};\\
&\textrm{at} \quad O(\lambda): \quad \epsilon^2\frac{dg}{dx} \left[ \textrm{Pe} \rho_0-\frac{\partial \rho_0}{\partial x} \right]+g \, \frac{\partial^2 \rho_0}{\partial z^2}+\frac{\partial \rho_1}{\partial z}=0 \quad \textrm{at} \quad z=\frac{1}{2}; \\
&\textrm{at} \quad O(\lambda ^2): \quad \epsilon^2\frac{dg}{dx}\left[\textrm{Pe}\left(\rho_1+g \, \frac{\partial \rho_0}{\partial z}\right)-g \, \frac{\partial^2 \rho_0}{\partial x \partial z}-\frac{dg}{dx}\frac{\partial \rho_0}{\partial z}-\frac{\partial \rho_1}{\partial x} \right]\\
& \qquad \qquad \qquad+\frac{\partial \rho_2}{\partial z}+g \, \frac{\partial^2 \rho_1}{\partial z^2} + \frac{g^2}{2} \, \frac{\partial^3 \rho_0}{\partial z^3}=0 \quad \textrm{at} \quad z=\frac{1}{2}.
\end{split}
\end{equation}

Finally, we also need to transform the normalization condition in the entire channel to an asymptotically equivalent condition for a domain limited by the two parallel planes defined by $z=\pm1/2$.
First, we substitute the regular perturbation
expansion for the asymptotic probability distribution into the normalization condition,
\begin{equation}
\Phi(\lambda)=\int_0^1 dx\int_{-1/2-\lambda g(x)}^{1/2+\lambda g(x)}\left(\rho_0+\lambda \rho_1+\lambda^2 \rho_2+O(\lambda^3) \right)dz=1.
\end{equation}
Then, we expand the normalization integral $\Phi(\lambda)$ in a Taylor series in the neighborhood of $\lambda=0$,
\begin{equation}
\Phi(\lambda)=\Phi(0)+\lambda\frac{d\Phi(0)}{d\lambda}+\frac{1}{2}\lambda^2\frac{d^2\Phi(0)}{d\lambda^2}
+O(\lambda^3).
\end{equation}
Calculating the expansion above explicitly, substituting it in the normalization condition, and equating like powers of $\lambda$, we obtain the equivalent conditions
on the distribution between the two parallel planes:
\begin{equation}
\begin{split}
&\textrm{at}\quad O(1):\quad \int_0^1dx\int_{-1/2}^{1/2}\rho_0 dz=1;\\
&\textrm{at}\quad O(\lambda):\quad \int_0^1\left(\rho_0\big|_{z=1/2}+\rho_0\big|_{z=-1/2}\right)g \, dx + \int_0^1dx\int_{-1/2}^{1/2}\rho_1dz=0; \\
&\textrm{at}\quad O(\lambda^2): \quad \frac{1}{2}\int_0^1\left[\left(g \, \frac{\partial \rho_0}{\partial z}+2\rho_1\right)\bigg|_{z=1/2}+\left(g \, \frac{\partial \rho_0}{\partial z}+2\rho_1\right)\bigg|_{z=-1/2}\right]g \,dx \\
&\hspace{3cm}+ \int_0^1dx\int_{-1/2}^{1/2}\rho_2dz=0.
\end{split}
\end{equation}

%%%%%%%%%%%%%%%%%%%%%%%%%%%%%%%%%%%%%%%%%%%%%%%%%%%%%%%%%%%%%%%%%%%%%%%%%%%%%%%%%%%%
\section{Leading order correction in the solid channel}
%%%%%%%%%%%%%%%%%%%%%%%%%%%%%%%%%%%%%%%%%%%%%%%%%%%%%%%%%%%%%%%%%%%%%%%%%%%%%%%%%%%%
\label{app_p1}
We seek a solution of Eq. (\ref{eqn:gov-solid}) with the corresponding boundary and no-flux conditions to $O(\lambda)$.
Given the linearity of the equation and the boundary and normalization conditions, we propose two base solutions of the form
\begin{equation}
\rho_1(x,z)=f_1(z)e^{2\pi ix} \quad\textrm{and}\quad g_1(z)e^{-2\pi i x}.
\label{eqn:p1_solid}
\end{equation}
Substituting these proposed solutions into Eq. (\ref{eqn:gov-solid}) we obtain
\begin{eqnarray}
f_1(z)= A_{f1} \cosh(\alpha_1 z),\\
g_1(z)=A_{g1} \cosh(\overline{\alpha_1}z),
\end{eqnarray}
where $\alpha_1$ is the complex number
\begin{equation}
\alpha_1^2 = \epsilon^2(2\pi i \textrm{Pe}+4\pi^2),
\end{equation}
and $\overline{\alpha_1}$ is its complex conjugate.
Since the probability distribution is a real value function we simplify the problem assuming $A_1=A_{f1}=\overline{A_{g1}}$,
\begin{equation}
\rho_1(x,z)=A_1 \cosh(\alpha_1 z)e^{2\pi i x}+\overline{A_1 \cosh(\alpha_1 z)}e^{-2\pi i x}.
\end{equation}
This proposed solution clearly satisfies the periodicity condition in $x$, as well as the normalization condition. It is also symmetric about the $z-$axis, by construction. Therefore, the only remaining condition is the
no-flux condition. Replacing then the proposed solution into Eq. (\ref{eqn:asy_flux}) we determine the constant $A_1$,
\begin{equation}
A_1=-\frac{\pi \epsilon^2\textrm{Pe}}{\alpha_1 \sinh(\alpha_1/2)}.
\end{equation}

%Merlin.mbs v4.21 2009-07-09.
%

%\bibliography{asymptotic}
%\bibliography{d:/JHU_research/asymptotic}
\end{document}